\documentclass[a4paper,twocolumn,showpacs,nofootinbib,prc]{revtex4}
\usepackage{graphicx}
\usepackage{amsfonts}
\usepackage{amsmath}
\usepackage{amsbsy}
\renewcommand{\vec}[1]{\mathbf{#1}}
\def\be{\begin{equation}}
\def\ee{\end{equation}}
\def\bea{\begin{eqnarray}}
\def\eea{\end{eqnarray}}
\newcommand{\beq}{\begin{equation}}
\newcommand{\eeq}[1]{\label{#1} \end{equation}}
\def\Kappa{\mathcal{K}}

\begin{document}

\title{Covariant description of kinetic freeze out through a finite space-like layer}

\author{
E. Moln\'ar$^1$, L. P. Csernai$^{1,2}$, V. K. Magas$^{1,3}$,
A. Ny\'iri$^1$ and K. Tamosiunas$^1$}


\affiliation{$^1$ Section for Theoretical and Computational
Physics, and Bergen Computational Physics Laboratory, BCCS-Unifob,
University of Bergen, Allegaten 55, 5007 Bergen, Norway\\
$^2$ MTA-KFKI, Research Institute of Particle and Nuclear Physics,\\
H-1525 Budapest 114, P.O.Box 49, Hungary\\
$^3$ Departamento de F\'{\i}sica Te\'orica and IFIC Centro Mixto\\
Universidad de Valencia-CSIC, Institutos de Investigaci\'on de Paterna \\
Apdo. correos 22085, 46071, Valencia, Spain}

\begin{abstract}
{
Abstract: The problem of Freeze Out (FO) in relativistic heavy ion reactions is addressed.
We develop and analyze an idealized one-dimensional model of FO in a finite layer, based
on the covariant FO probability.
The resulting post FO phase-space distributions are discussed for different FO probabilities
and layer thicknesses.
}
\end{abstract}

\pacs{24.10.Nz, 25.75.-q}

\maketitle
\section{Introduction}

The hydrodynamical description of relativistic particle collisions was first discussed more
than 50 years ago by Landau \cite{Landau_1} and nowadays it is frequently used in different
versions for simulations of heavy ion collisions.
Such a simulation basically includes three main stages.
The initial stage, the fluid-dynamical stage and the so-called Freeze Out (FO) stage when
the hydrodynamical description breaks down.
During this latter stage, the matter becomes dilute, cold, and non-interacting, the particles
stream toward the detectors freely, their momentum distribution freezes out.
Thus, the freeze out stage is essentially the last part of a collision process and the
source of the observables.
\\ \indent
The usual recipe is to assume the validity of hydrodynamical treatment up to a sharp FO
hypersurface, e.g. when the temperature reaches a certain value, $T_{FO}$.
When we reach this hypersurface, all interactions cease and the distribution of particles
can be calculated.
\\ \indent
In such a treatment FO is a discontinuity where the properties of the matter change suddenly
across some hypersurface in space-time.
The general theory of discontinuities in relativistic flow was first discussed by Taub \cite{Taub}.
That description can only be applied to discontinuities across propagating hypersurfaces, which
have space-like, $(d\sigma_{\mu} d\sigma^{\mu} = -1)$, normal vector.
The discontinuities across hypersurfaces with time-like, $(d\sigma_{\mu} d\sigma^{\mu} = 1)$,
normal vector were considered unphysical.
The remedy for this came only 40 years later in \cite{Csernai_87}, generalizing Taub's approach
for both time-like and space-like hypersurfaces.
Consequently, it is possible to take into account conservation laws exactly across any surface of
discontinuity in relativistic flow.
\\ \indent
As it was shown recently in \cite{Bugaev_1,cikk_0}, the frequently used Cooper-Frye
prescription \cite{Cooper-Frye} to calculate post FO particle spectra gives correct
results only for discontinuities across time-like normal vectors.
The problem of negative contributions in the Cooper-Frye formula was healed by a simple
cut-off, $\Theta(p^{\mu}d\sigma^{\mu})$, proposed by Bugaev \cite{Bugaev_1}.
However, this formulation is still based on the existence of a sharp FO hypersurface, which
is a strong idealization of a FO layer of finite thickness \cite{cikk_2}.
Thus, by assuming an immediate sharp FO process, the questions of final state interactions
and the departure from local equilibrium are left unjustified.
\\ \indent
The recent paper \cite{ModifiedBTE,ModifiedBTE2} formulates the freeze out problem in the framework
of kinetic transport theory.
The dynamical FO description has to be based on the Modified Boltzmann Transport Equation (MBTE),
rather than on the commonly used Boltzmann Transport Equation (BTE).
The MBTE abandons the local molecular chaos assumption and the requirement of smooth variation of
the phase-space distribution, $f(x,p)$, in space-time.
This modification of BTE, makes it even more difficult to solve the FO problem from first principles.
Therefore, it is very important to build phenomenological models, which can explain the basic
features of the FO process.
\\ \indent
The present paper aims to build such a simple phenomenological model.
The kinetic approach presented, is applicable for FO in a layer of finite thickness
with a space-like normal vector.
It can be viewed as a continuation and generalization of \cite{cikk_1, cikk_3,cikk_4}.
The kinetic model for FO in time-like direction was discussed in a recent paper \cite{cikk_5},
however, the fully covariant model analysis and the treatment are presented in \cite{article_2}.
\\ \indent
In present work we use stationary, one-dimensional FO models for the transparent presentation.
Such models can be solved semianalytically, what allows us to trace the effects of
different model components, assumptions and restrictions applied on the FO description.
We do not aim to apply directly the results presented here to experimental heavy ion collision data,
instead our purpose is to study qualitatively the basic features of the FO process.
We want to demonstrate the applicability of the proposed covariant FO escape rate,
and most importantly, to see the consequences of finishing FO in a finite layer.
Up to now, two extreme ways of describing FO were used:
(i) FO on an infinitely narrow hypersurface and
(ii) infinitely long FO in volume emission type of model.
To our knowledge this is the first attempt to, at least qualitatively, understand how FO in finite
space-time domain can be simulated and what will be its outcome.
In such stationary, one-dimensional models the expansion cannot be realistically included, therefore it
is ignored.
\\ \indent
In realistic simulations of high energy heavy-ion reactions the full 3D description of expanding
and freezing out system should be included.
This work is under initial development.

\section{Freeze out from kinetic theory }\label{FO_distribution}

A kinetic theory describes the time evolution of a single particle distribution function,
$f(x,p) = f(t,\vec{x},p^0,\vec{p})$, in the 6D phase-space.
To describe freeze out in a kinetic model, we split the distribution function into two
parts \cite{cikk_1,Grassi,Sinyukov}:
\be
f(x,p) = f_i (x,p) + f_f (x,p) \, .
\ee
The free component, $f_f$, is the distribution of the frozen out particles, while
$f_i$, is the distribution of the interacting particles.
Initially, we have only the interacting part, then as a consequence of FO dynamics,
$f_i$ gradually disappears, while $f_f$ gradually builds up.
In this paper we convert the description of the FO process from a sudden FO,  i.e. on a sharp
hypersurface, into a gradual FO, i.e. in some finite space-time domain.
\\ \indent
Freeze out is known to be a strongly directed process \cite{kemer}, where the particles are allowed to cross
the FO layer only outwards, in the direction of the normal vector, $d\sigma_{\mu}$, of the FO hypersurface.
Many dynamical processes happen in a way, where the phenomenon propagates into some direction, such
as detonations, deflagrations, shocks, condensation waves, etc.
Basically, this means that the gradients of the described quantity (the distribution function in our case)
in all perpendicular directions can be neglected compared to the gradient in the given
direction $d \sigma_{\mu}$, i.e. $\vec{\bigtriangledown} f \approx d \sigma^{\mu} \partial_\mu f$.
In such a situation these can be effectively described as one-dimensional processes,
and the space-time domain, where such a process takes place, can be viewed as a layer.
\\ \indent
Therefore, we develop a one-dimensional model for the FO process in a layer of finite thickness, $L$.
We assume that the boundaries of this layer are approximately parallel, and thus, the
thickness of the layer does not vary much.
This can be justified, for example, in the case when the system size is much larger than $L$.
At the inside boundary of this layer there are only interacting particles, whereas at the outside
boundary all particles are frozen out and no interacting particles remain.
Note that the normal to the FO layer, $d\sigma^{\mu}$, can be both space-like or time-like.
\\ \indent
The gradual FO model for the infinitely long one-dimensional FO process was presented in recent works
\cite{cikk_1,cikk_3,cikk_4}.
We are going to build a similar model, but now we make sure that FO is completely finished within
a finite layer.

\subsection{Freeze out in a finite layer}\label{Fid}

\begin{figure}[!htb]
\centering
\includegraphics[width=7.4cm, height = 6.4cm]{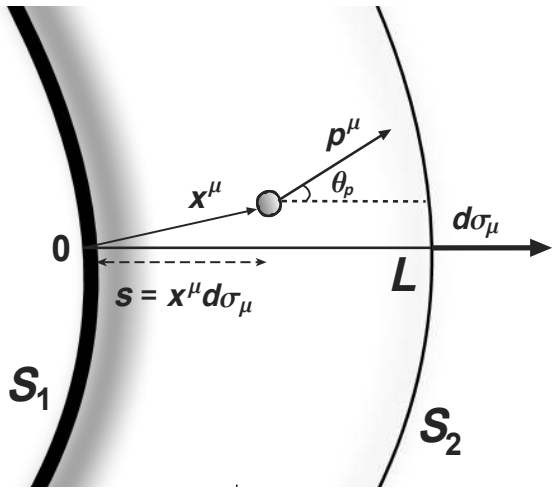}
\caption{ The picture of a gradual FO process within the finite FO layer, in x-direction, i.e.
$d\sigma_{\mu}=(0,1,0,0)$.
The particles are moving in different directions outwards, indicated by the angle $\theta$ .
The inside boundary of the FO layer, $S_1$ (thick line) indicates the points
where the FO starts.
This is the origin of the coordinate vector, $x^{\mu}$.
Within the finite thickness of the FO layer, $L$, the density of the
interacting particles gradually decreases (indicated by shading) and disappears
at the outside boundary, $S_2$ (thin line) of the FO layer.
}\label{gradualFO}
\end{figure}

In kinetic theory the interaction between particles is due to collisions.
A quantitative characterization of collisions is given by the mean free path (m.f.p.),
giving the average distance between collisions.
The m.f.p., $\lambda_{mfp}$, is inversely proportional to the density, $\lambda_{mfp} \sim 1/n(x)$.
If we have a finite FO layer, the interacting particles inside this domain must have a finite m.f.p.
During the FO process, as the density of the interacting particles decreases, they are entering
into a collisionless regime, where their final m.f.p., tends to infinity, or at least,
gets much larger than the system size $L$.
The realistic FO process for nucleons in a heavy ion collisions happens within a finite
space-time FO domain, which has a thickness of a few initial mean free paths \cite{Csernai_Bravina}.
Hence, one must realize that the FO process cannot be fully exploited by the means of the m.f.p. concept,
since we have to describe a process where we have on average a few collisions per particle before freeze out.
Therefore, this type of processes should be analyzed by having also another characteristic length scale
different from the m.f.p.
In our case it should be related to the thickness, $L$, of the FO layer.
\\ \indent
Recent conjectures based on strong flow and relatively small dissipation find that the state where
collective flow starts is strongly interacting and strongly correlated while the viscosity
is not large \cite{Csernai_hydro_2005}.
This indicates a small m.f.p., in the interacting matter, while at the surface $\lambda_{mfp} \rightarrow \infty$.
Several indications point out that in high energy heavy ion reactions freeze out and hadronization
happens simultaneously from a supercooled plasma \cite{Csorgo_Csernai,Csernai_Mishustin,Keranen}.
This could be modeled in a way that pre-hadron formation and clusterization starts gradually in the plasma,
and this process is coupled to FO in a finite layer.
The FO is finished when the temperature of the interacting phase drops under a critical
value and all quarks cluster into hadrons, which no longer collide.
This is the possible qualitative scenario with well defined finite thickness $L$ of the FO layer.
\\ \indent
Now, let us recall the equations describing the evolution in the simple kinetic
FO model \cite{cikk_1, cikk_3,cikk_4}.
Starting from a fully equilibrated J\"uttner distribution, $f_{J}(p)$, i.e. $f_{i}(s=0,p) = f_{J}(p)$
and $f_{f}(s=0,p) = 0$, the two components of the momentum distribution develop in the direction of
the freeze out, i.e. along $d \sigma^{\mu}$, according to the following differential equations:
\bea\label{first} \nonumber
\partial_s f_{i} (s,p) ds &=& - f_{i}(s,p) W_{esc} (s,p) \, , \\
\partial_s f_{f} (s,p) ds &=& +  f_{i}(s,p) W_{esc} (s,p)\, ,
\eea
where $W_{esc} (s,p)$ is the escape rate governing the FO development and $s = x^{\mu} d\sigma_{\mu}$.
Here $x^{\mu}$ is a 4-vector having its
origin\footnote{Any point of the inner surface, $S_1$, can be considered as an origin,
since translations along $S_1$ do not change $s$, the projection of $x^\mu$ on the FO normal vector,
$d \sigma^\mu$, as long as $S_1$ and $S_2$ are parallel, as assumed. Of course, this latter assumption
can be justified only locally, in some finite region, as it is clear from Fig. \ref{gradualFO}.}
at the inner surface, $S_1$, of the FO layer, see Fig. \ref{gradualFO}.
In order to obtain the probability to escape, for a particle passing from $0$ till $s$, $\mathcal{P}_{esc}(s)$,
we have to integrate the escape rate along a trajectory crossing the FO layer:
\beq
f_{i}(s,p) = f_J(p)
\underbrace{\exp \! \left( \!- \! \int_{0}^{s} \! ds' W_{esc} (s',p) \right)}_{1 - \mathcal{P}_{esc}} \, .
\eeq{assmptotic_condition}
The definition for the escape probability was previously given in \cite{Sinyukov}, in terms of collision
or scattering rates, where the FO process was lasting infinitely long.
In our finite layer FO description the quantity that defines the escape probability is the escape rate.
\\ \indent
To have a complete physical FO finished at a finite distance/time, we require:
$\mathcal{P}_{esc} \rightarrow 1$, when $s\rightarrow L$.
In usual cascade models the probability of collision never becomes exactly zero,
and correspondingly $\mathcal{P}_{esc}$ never becomes exactly one, and the FO process lasts ad infinitum.
This is due to the fact that the probability of collision is calculated based on the thermally
averaged cross section, which does not vanish for thermal, e.g. Gaussian, momentum distributions.
In reality the free or frozen out particles have no isotropic thermal distributions but these distributions
can be anisotropic and strongly confined in the phase-space.
This means that the collision probability can be exactly zero and FO may be completed in a finite
space-time domain.
\\ \indent
It seems reasonable to parameterize the escape rate, which has dimension
one over length, in terms of some characteristic FO length, $\lambda'(s,p)$,
\beq
W_{esc} (x^{\mu},p^{\mu},d\sigma_{\mu}) \equiv \frac{1}{\lambda'(s,p)} \Theta(p^s)\, ,
\eeq{esca}
where the cut-off factor, $\Theta(p^s)\equiv \Theta(p^{\mu}d \sigma_{\mu})$, forbids the FO of particles
with momenta not pointing outward \cite{Bugaev_1}.
This FO parameter, $\lambda'(s,p)$, is not necessarily an average distance in space or duration in
time between two subsequent collisions, like the m.f.p.
The m.f.p., tends to infinity as the density decreases, while the FO just becomes faster in this limit.
Actually, the FO scale behaves in the opposite way to the m.f.p.
This can be seen, for example, in a simple, purely geometrical freeze out model, which takes into account
the divergence of the flow in a 3D expansion \cite{Bondorf}.
Both this and the phase transition or clusterization effect described at the beginning of this section, lead
to a finite FO layer $L$, even if the m.f.p., $\lambda_{mfp} \cong 1/n\sigma$ is still finite at the outer
edge of this layer.
\\ \indent
We consider the thickness of the layer $L$ to be the "proper" thickness of the FO layer,
because it depends only on invariant scalar matter properties like cross section, proper density,
velocity divergence, phase transition or clusterization rates.
These should be evaluated in the Local Rest frame (LR) of the matter, and since the layer is finite,
around the middle of this layer.
The proper thickness is analogous to the proper time, i.e. time measured in the rest frame of
the particle, hence the proper thickness is the thickness of the FO layer measured in the rest
frame attached to the freeze out front, that is, the Rest Frame of the Front (RFF).
Some of the parameters like the velocity divergence and the phase transition rate describe the dynamical changes
in the layer, so these can determine the properties, e.g. the thickness, of the finite layer.
However, calculating $L$ from the above mentioned properties is beyond the scope of this paper,
and $L$ is treated as a parameter in the following.
\\ \indent
Let us consider the Rest Frame of the Front, where the normal vector of the front points
either in time, $t$, or in space, $x$, direction, introducing the following
notations\footnote{
\qquad \qquad \qquad \qquad  Time-like \quad  \qquad Space-like  \\
$d\sigma_{\mu}$  \qquad \qquad \qquad  $(1,0,0,0) $ \qquad \quad $(0,1,0,0)$ \\
$s \, \, \, \equiv \, (d\sigma_{\rho} x^{\rho} )  $ \qquad \qquad $ t $  \qquad \qquad \qquad $ x  $ \\
$ p^s  \equiv \, (d\sigma_{\rho} p^{\rho} ) $ \qquad  \qquad  $ p^0 $  \qquad  \qquad \quad \, $ p^x $ \\
$ \partial_{s} \equiv \, (d\sigma^{\rho} \partial_{\rho}) $ \qquad \qquad  $\partial_{t}$ \qquad
\qquad \quad \, $\partial_{x} $ \\
$ \lambda'(s) \,$  \qquad\quad \qquad \qquad $\tau'(t)$  \qquad \qquad \, $\lambda'(x)$
}.
Indeed, if $d\sigma_{\mu}$ is space-like the resulting equations can be transformed into a
frame where the process is stationary (here $d\sigma_{\mu}=(0,1,0,0)$ and correspondingly $s\equiv x$),
while in the case of a time-like normal vector the equations can be transformed into a frame
where the process is uniform and time-dependent ($d\sigma_{\mu}=(1,0,0,0)$, $s\equiv t$).
For sake of transparency and simplicity we will perform calculations only for FO in a finite layer
with space-like normal vector in this paper, but many intermediate
results can and will be obtained in Lorentz invariant way.
\\ \indent
Inside the FO layer particles are separated into still colliding or interacting and
not-colliding or free particles.
The probability not to collide with anything on the way out, should depend on the number
of particles, which are in the way of a particle moving outward in the direction $\vec{p}/|\vec{p}|$
across the FO layer of thickness $L$, see Fig. \ref{gradualFO}.
If we follow a particle moving outward form the beginning, ($x^{\mu}=0$), i.e. the inner surface
of the FO layer, $S_1$, to a position $x^\mu$, there is still a distance
\be\nonumber
\frac{L - s}{\cos \theta_{\vec{p}} }
\ee
ahead of us, where $\theta_{\vec{p}}$ is the angle between the normal vector and $\vec{p}/|\vec{p}|$.
As this remaining distance becomes smaller the probability to freeze out becomes larger,
thus, we may assume that the escape rate is inversely proportional to some power, $a$, of this
quantity \cite{magastalk1, ModifiedBTE2}.
\\ \indent
Based on the above assumptions we write the escape rate as:
\beq
W_{esc} = \frac{1}{\lambda} \left( \frac{L}{L - s} \right)^a \left(\cos \theta_{\vec{p}}\right)^a \Theta(p^s)\, ,
\eeq{esc_b}
where this newly introduced parameter, $\lambda$, is the initial, i.e. at $S_1$,
characteristic FO length of the interacting matter, $\lambda=\lambda'(s=0,\cos \theta_{\vec{p}}=1)$.
The power $a$ is influencing the FO profile across the front.
Indeed, calculating the escape probability, $\mathcal{P}_{esc}$, eq. (\ref{assmptotic_condition}),
with the escape rate, given by eq. (\ref{esc_b}), we find
\be\nonumber
\mathcal{P}_{esc} = 1 - \left( \frac{L-s}{L}\right)^{\frac{L}{\lambda} \cos \theta_{\vec{p}} \Theta(p^s)} \, ,
\label{a1}
\ee
for $a=1$, and
\be\nonumber
\mathcal{P}_{esc} = 1- \exp \!\left[ \frac{L}{\lambda}\, \Theta(p^s) \frac{(\cos\theta_{\vec{p}} )^a}{(a-1)}
\! \left(\! 1 - \left(\frac{L}{L-s}\right)^{a-1}\! \right) \! \right]\, ,
\label{a2}
\ee
for $a\neq 1$.
Thus, we see that for different $a$-values we have different FO profiles:
\begin{itemize}
\item [] $a=1$: power like FO,
\item [] $a>1$: fast, exponential like FO,
\item [] $a<1$: no complete FO within the finite layer, since $\mathcal{P}_{esc}$ does not tend to $1$ as
$s$ approaches  $L$.
\end{itemize}
In papers \cite{cikk_1,cikk_3,cikk_4} the authors were using $a=1$, and were modeling FO in an infinite layer.
In order to study the effects of FO within a finite space-time domain, we would like
to compare the results of our calculations with those of earlier works,
therefore we shall also take $a=1$ in further calculations.
It is easy to check that our escape rate, eq. (\ref{esc_b}), equals the earlier expression
\be \nonumber
P_{esc}=\frac{\cos \theta_{\vec{p}}}{\lambda}\,  \Theta(p^s)
\ee
in $L\rightarrow \infty$ limit.
Thus, the model discussed in this paper is a generalization of the models for infinitely long FO, described
in \cite{cikk_1,cikk_3,cikk_4}, and allows us to study FO in a layer of finite thickness.
\\ \indent
The angular factor, $\cos \theta_{\vec{p}}$, maximized the FO probability for those particles, which propagate
in the direction closest to the normal of the FO layer.
For the FO in time-like directions, studied in \cite{cikk_5}, the angular factor was $1$.
This factor, and correspondingly the escape rate, eq. (\ref{esc_b}), are not covariant.
Furthermore, this earlier formulation does not take into account either that the escape
rate of particles should be proportional to the particle velocity
(the conventional non-relativistic limit of the collision rate
contains the thermal average $<\sigma v>$).
Let us consider the simplest situation, when the Rest Frame of the Front is the same as Rest
Frame of the Gas (RFG), where the flow velocity is $u^{\mu}=(1,0,0,0)$.
If freeze out propagates in space-like direction, i.e.  $d\sigma_{\mu} = (0,1,0,0)$,
as shown in Fig. \ref{gradualFO}, then $\cos \theta_{\vec{p}} = p^{x}/|\vec{p}|$.
Therefore, a straightforward generalization of the escape rate, based on the above arguments, is
\beq
\cos \theta_{\vec{p}} \times |\vec{v}|
\equiv \frac{p^{x}}{|\vec{p}|}\times\frac{|\vec{p}|}{p^{0}}
=\bigg( \frac {p^\mu d\sigma_\mu}{p^\mu u_\mu} \bigg) \, ,
\eeq{angular}
where the r.h.s. of this equation is an invariant scalar in covariant form.
Now, we assume that this simple generalization is valid for any space-like or
time-like FO direction, even when RFG and RFF are different \cite{kemer,QM05etele,ModifiedBTE2}.
\\ \indent
Based on the above arguments, we can write the total escape rate from eq. (\ref{esc_b})
in a Lorentz invariant\footnote{Now, it is important that $L$ is defined as an invariant scalar, so $W_{esc}$
is also an invariant scalar.}
form:
\be\label{esc1}
W_{esc} = \frac{1}{\lambda} \left( \! \frac{L}{L - s} \! \right)
\! \! \left( \frac{p^s}{p^{\mu}u_{\mu}} \right) \! \Theta(p^s)\, ,
\ee
which now opens room for general study of FO in relativistic flow in layers of any thickness.
\\ \indent
Former FO calculations in \cite{cikk_1, cikk_3,cikk_4} were always performed in RFF.
Aiming for semianalytical results and transparent presentation, as well as in
order to compare our results with former calculations, we will also study the system evolution in RFF,
but now this is only our preference.
In principle calculations can be performed in any reference frame.
In more realistic many dimensional models, which will take into account the system expansion simultaneous
with the gradual FO, it will be probably more adequate to work in RFG or in Lab frame, and our
invariant escape rate, eq. (\ref{esc1}), can be directly used as a basic FO ingredient of such models.

\subsection{The Lorentz invariant escape rate}

\begin{figure}[!htb]
\centering
\includegraphics[width=8.4cm, height = 6.8cm]{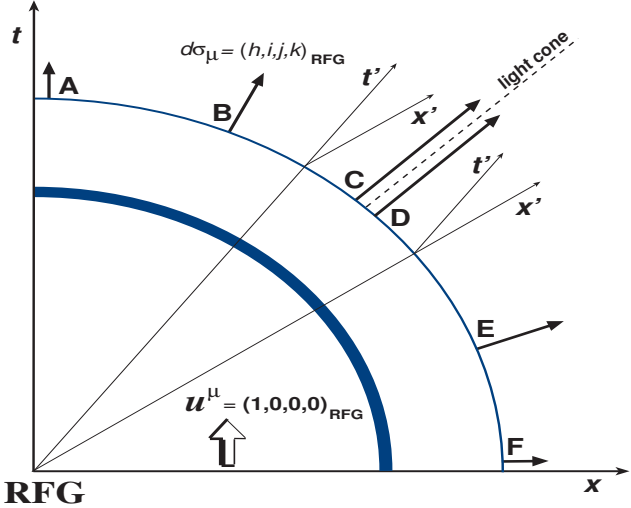}
\caption{A simple FO hypersurface in the Rest Frame of the Gas (RFG:\,[t,x]),
where $u^{\mu} = (1,0,0,0)_{RFG}$, including time-like and space-like parts.
The normal vector of the FO front, $d\sigma_{\mu}$, is a time-like 4-vector
at the time-like part and it is changing smoothly into a space-like 4-vector
in the space-like part.
On these two parts of the hypersurface, in the Rest Frame of the Front (RFF),
$d\sigma_{\mu}$ points into the direction of the $t'$ ($x'$)-axis respectively.
At points A,\, B,\, C,\, D,\, E,\, F, we have different Rest Frame(s) of the Front, (RFF:\,[t',x']).
}
\label{updatedFO}
\end{figure}

\begin{figure}[!htb]
\centering
\includegraphics[width=8.6cm, height = 6.8cm]{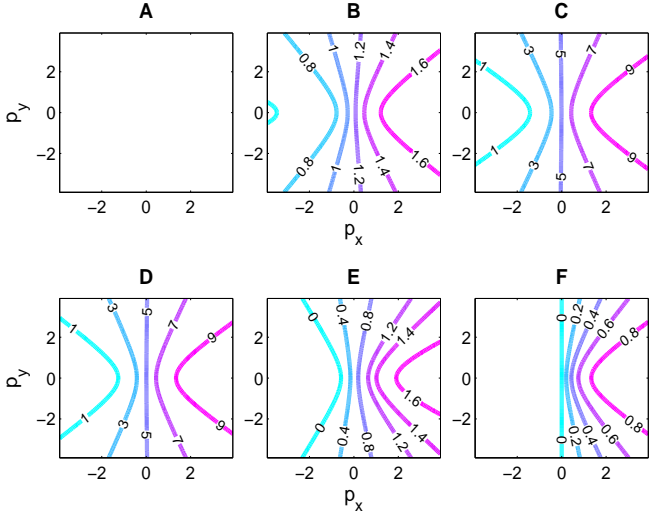}
\caption{The contour plots of the momentum dependent part of the escape
rate, $W(p)$, as in eq. (\ref{esc_probability_draw}), presented in six subplots
at different points of the FO hypersurface.
All plots are in RFG.
For region $A$: $d{\sigma}_\mu = (1,0,0,0)$, and $W(p)$ is one uniformly, for $B$: $d{\sigma}_\mu = \gamma_{\sigma}
(1,0.5,0,0)$, $C$: $d{\sigma}_\mu = \gamma_{\epsilon}(1.01,0.99,0,0)$, (first row),
$D$: $d{\sigma}_\mu = \gamma_{\epsilon}(0.99,1.01,0,0)$,
$E$: $d{\sigma}_\mu = \gamma_{\sigma} (0.5,1,0,0)$, $F$: $d{\sigma}_\mu = (0,1,0,0)$,
(second row). The momenta are in units of particle mass, [m].}
\label{P_6_RFG}
\end{figure}

\begin{figure}[!hbt]
\centering
\includegraphics[width=8.6cm, height = 6.8cm]{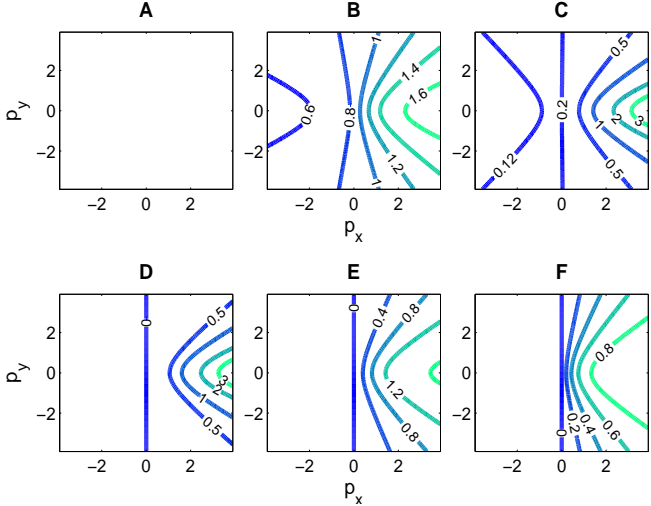}
\caption{The contour plots of the momentum dependent part of the escape
rate, $W(p)$, in RFF.
For region $A$: $u_\mu = (1,0,0,0)$, and $W(p)$ is one uniformly, for $B$: $u_\mu = \gamma_{\sigma}
(1,-0.5,0,0)$, $C$: $u_\mu = \gamma_{\epsilon}(1.01,-0.99,0,0)$, (first row),
$D$: $u_\mu = \gamma_{\epsilon}(1.01,-0.99,0,0)$,
$E$: $u_\mu = \gamma_{\sigma} (1,-0.5,0,0)$, $F$: $u_\mu = (1,0,0,0)$,
(second row). In cases D, E, F the escape rate vanishes for momenta with $p_x<0$.
The momenta are in units of particle mass, [m].}
\label{P_6_RFF}
\end{figure}

In this section let us study this new angular factor, in more detail.
We will take the $p$-dependent part of the escape rate, eq. (\ref{esc1}), and denote it as:
\beq
W(p) = \frac{p^{\mu}d\sigma_{\mu}}{p^{\mu}u_{\mu}} \,
\Theta(p^{\mu}d\sigma_{\mu}) \, .
\eeq{Wp}
In RFG, where the flow velocity of the
matter is $u^{\mu}=(1,0,0,0)_{RFG}$ by definition, $W(p)$, is given as
\be \label{esc_probability_draw}
W(p) = \frac{p^{\mu}d\sigma_{\mu}}{p^{0}} \, \Theta(p^{\mu}d\sigma_{\mu}) \, \bigg{|}_{RFG} \, ,
\ee
and it is smoothly changing as the direction of the normal vector changes in RFG.
This will be discussed in more detail in the rest of this section.
\\ \indent
In the following, we will take different typical points of
the FO hypersurface, A, B, C, D, E, F, see Fig. \ref{updatedFO}.
At these points, the normal vectors of the hypersurface, $d\sigma_{\mu} = (h,i,j,k)_{RFG}$, are given below.
\\
\\ \indent
To calculate the normal vector for different cases shown in Fig. \ref{updatedFO},
we simply make use of the Lorentz transformation.
The normal vector of the time-like part of the FO hypersurface may be defined as the local
$t'$-axis, while the normal vector of the space-like part may be defined as the local $x'$-axis.
As $d{\sigma}_\mu$ is normalized to unity, its components may be interpreted in terms
of $\gamma_{\sigma}$ and $v_{\sigma}$, where $\gamma_{\sigma} = \frac{1}{\sqrt{1 - v_{\sigma}^2}}$.
So, we have:
\begin{itemize}
\item [A)] \  $d\sigma_{\mu} = (1,0,0,0)$, leads to $W(p) = 1$,

\item [B)] \  $d\sigma_{\mu} = \gamma_{\sigma} (1,v_{\sigma},0,0)$,  leads to
                $W(p) = \frac{\gamma_{\sigma} ( p^0 + v_{\sigma} p^x)}{p^0}$,

\item [C)] \  $d\sigma_{\mu} = \gamma_{\epsilon} (1+\epsilon,1-\epsilon,0,0)$, where
              $\gamma_{\epsilon} = (4\epsilon)^{-\frac{1}{2}}$, $\epsilon \ll 1$.
              This leads to
              $W(p) = \gamma_{\epsilon} \frac{(p^0 + p^x) + \epsilon (p^0 - p^x)}{ p^0}  $,

\item [D)] \  $d\sigma_{\mu} = \gamma_{\epsilon} (1-\epsilon,1+\epsilon,0,0)$,  where
              $\gamma_{\epsilon} = (4\epsilon)^{-\frac{1}{2}}$, \\ $\epsilon \ll 1$. This leads to
                $W(p)=\gamma_{\epsilon} \frac{(p^0 + p^x) - \epsilon (p^0 - p^x)}{ p^0}\\
                \times \Theta \left( \gamma_{\epsilon}(p^0 + p^x) - \gamma_{\epsilon} \epsilon (p^0 - p^x)\right) $,

\item [E)] \  $d\sigma_{\mu} = \gamma_{\sigma} (v_{\sigma},1,0,0)$, leads to \\
                $\quad \, W(p) = \frac{\gamma_{\sigma} (v_{\sigma} p^0 - p^x)}{p^0}
                \times  \Theta \Big( \gamma_{\sigma}(v_{\sigma} p^0 - p^x) \Big)$,

\item [F)] \  $d\sigma_{\mu} = (0,1,0,0)$, leads to  $W(p) = \frac{p^x}{p^0}
                \times  \Theta(p^{x})$.
\end{itemize}
\indent
The resulting phase-space escape rates are shown in Fig. \ref{P_6_RFG}
for the six cases described above.
\\ \indent
Similar calculations can be done in RFF, where $d\sigma_{\mu} = (1,0,0,0)$ for A, B, C and
$d\sigma_{\mu} = (0,1,0,0)$ for D, E, F, leading the following values for $W(p)$:
\begin{enumerate}
\item [A)] \  $u_{\mu} = (1,0,0,0)$, leads to $W(p) = 1$,

\item [B)] \  $u_{\mu} = \gamma_{\sigma} (1,-v_{\sigma},0,0)$,  leads to \\
              $W(p) = \frac{p^0}{\gamma_{\sigma} ( p^0 - v_{\sigma} p^x)}$,

\item [C)] \  $u_{\mu} = \gamma_{\epsilon} (1+\epsilon, -(1-\epsilon),0,0)$, where
              $\gamma_{\epsilon} = (4\epsilon)^{-\frac{1}{2}}$, $\epsilon \ll 1$.
              This leads to
              $W(p) = \frac{1}{\gamma_{\epsilon}} \frac{p^0} { (p^0 - p^x) \,+\, \epsilon (p^0 + p^x)}  $,

\item [D)] \  $u_{\mu} = \gamma_{\epsilon} (1+\epsilon, -(1-\epsilon),0,0)$, where
              $\gamma_{\epsilon} = (4\epsilon)^{-\frac{1}{2}}$, $\epsilon \ll 1$.
              This leads to
              $W(p) = \frac{1}{\gamma_{\epsilon}} \frac{p^0} { (p^0 - p^x) \,+\, \epsilon (p^0 + p^x)} \\
              \times  \Theta(p^x)$,

\item [E)] \  $u_{\mu} = \gamma_{\sigma} (1,-v_{\sigma},0,0)$, leads to \\
                $\quad \, W(p) = \frac{p^x}{\gamma_{\sigma} ( p^0 - v_{\sigma}p^x)}
                \times  \Theta ( p^x)$,

\item [F)] \  $u_{\mu} = (1,0,0,0)$, leads to  $W(p) = \frac{p^x}{p^0}
                \times  \Theta(p^{x})$.
\end{enumerate}
For these cases, A, B, C, D, E, F, in RFF the resulting
phase-space escape rates are shown in Fig. \ref{P_6_RFF}.
\\
\\ \indent
Figs. \ref{P_6_RFG} and \ref{P_6_RFF} show that the momentum dependence of the escape rate,
uniform in point A, becomes different at different points of the FO hypersurface, but this
change is continuous, when we are crossing the light cone, from point C to point D.
Although in RFF, Fig. \ref{P_6_RFF}, it seems that there is a principal difference between space-like
and time-like FO directions, due to the cut-off $\Theta(p^{\mu}d\sigma_{\mu})$ function, but
this is only the consequence of the chosen reference frame, i.e. RFF is defined in a way to stress
the difference between these two cases, since going from C to D, the normal vector has a jump, i.e.
$d\sigma_{\mu} = (1,0,0,0)$ goes over to $d\sigma_{\mu} = (0,1,0,0)$.
Nevertheless, $W(p)$ is a continuous function as we change $d\sigma_{\mu}$, and in other frames, for
example in RFG, Fig. \ref{P_6_RFG}, we can see this clearly.
\\

\subsection{The updated simple kinetic model}\label{Kinetic model}

Now, using the new invariant escape rate, eq. (\ref{esc1}), we can generalize the simple model presented
in \cite{cikk_1,cikk_3,cikk_4}, i.e. eqs. (\ref{first}), for a  finite space-time FO layer:
\bea \label{second} \nonumber
\partial_s f_{i}  \,ds &=& - \left( \! \frac{L}{L - s} \! \right) \!
\left(\frac{ p^{s}}{p^{\mu} u_{\mu}} \right) \!
\Theta(p^{s}) f_{i} \,\frac{ds}{\lambda} \, , \\
\partial_s f_{f}  \,ds &=& \left(\!\frac{L}{L - s}\! \right) \!
\left( \frac{p^{s}}{p^{\mu} u_{\mu}} \right)\!
\Theta(p^{s}) f_{i}\, \frac{ds}{\lambda} \, ,
\eea
Solving the first equation we find for the interacting component:
\bea\nonumber
f_{i} (s,p) &=& f_{J}(p) \left(\! \frac{L-s}{L}\right)^{\frac{L}{\lambda}
\left(\frac{p^{s}}{p^{\mu} u_{\mu}} \right)\Theta(p^{s})} \, \\
&\buildrel s\rightarrow L \over \longrightarrow & f_{J}(p) \, \Theta(-p^{s}) \, .
\eea
Now, inserting this result into the second differential equation, from eqs. (\ref{second}), we obtain the
FO solution, which describes the momentum distribution of the frozen out particles:
\bea \label{f_free}
\nonumber
f_{f}(s,p) &=& f_{J}(p) \,
\left[ 1 - \left( \!\frac{L-s}{L} \! \right)^{\frac{L}{\lambda}
\left( \frac{p^s}{p^{\mu} u_{\mu}} \right) \!
\Theta(p^s)} \right] \, \\
&\buildrel s\rightarrow L \over \longrightarrow & f_{J}(p) \, \Theta(p^s) \, .
\eea
As $s$ tends to $L$, i.e. to the outer boundary of the FO layer, this distribution, depending on the
direction of the normal vector (space-like) or time-like will tend to the (cut) J\"uttner distribution, $f_J(p)$.
This means that (part of) the original J\"uttner distribution survives even when we reach the outer
boundary of the FO surface.
To remedy this highly unrealistic result, in \cite{cikk_1,cikk_3,cikk_4,cikk_5}, rethermalization
in the interacting component was taken into account via the relaxation time approximation,
i.e. we insert into the equation for the interacting component a new term, which describes
that the interacting component approaches some equilibrated (J\"uttner) distribution, $f_{eq}(s)$,
with a relaxation length, $\lambda_0$:
\bea\label{first-rethermalized}
\partial_s f_{i}  ds &=& - \left(\! \frac{L}{L-s}\! \right) \!
\left(\frac{ p^s}{p^{\mu} u_{\mu}} \right) \!
\Theta(p^s) f_{i} \, \frac{ds}{\lambda} \\ \nonumber
&+& \quad \left[ f_{eq}(s) - f_{i} \right] \frac{ds}{\lambda_0} \, , \\
\partial_s f_{f}  ds &=&  \left(\! \frac{L}{L - s}\! \right)\!
\left(\frac{ p^s}{p^{\mu} u_{\mu}} \right) \!
\Theta(p^s) f_{i} \, \frac{ds}{\lambda} \, .
\label{second-rethermalized}
\eea
Let us concentrate on the equation for the interacting component.
Here the first term from eq. (\ref{first-rethermalized}), related to FO, moves the distribution out
of the equilibrium, and decreases the energy-momentum density and baryon density of the interacting particles.
The second term from eq. (\ref{first-rethermalized}), changes the distribution in the direction
of the thermalization, while it does not effect the conserved quantities.
The relative strength of the FO and rethermalization processes is determined by the two characteristic
lengths, $\lambda$ and $\lambda_0$.
\\ \indent
In general the evolution of the interacting component can be solved numerically or
semianalytically, at every step of the integration.
Then, the change of conserved quantities due to FO should be evaluated using the actual
distribution, $f_i (s,p)$ at the corresponding point $s$.
For the purpose of this work, namely for the qualitative study of the FO features, it is enough to
use an approximate solution, similarly as it was done in  \cite{cikk_3,cikk_4,cikk_5}.
This would also allow us to make a direct comparison with results of these older calculations.
Thus, the evaluation of the change of the conserved quantities is done analytically, i.e. $f_i (s,p)$
is approximated with an equilibrium distribution function $f_{eq}(s)$ with
parameters, $T(s), \, n(s), \, u^{\mu}(s)$.
\\ \indent
This approximation is based on the fact that in most physical situations the overall number of particle
collisions vastly exceeds the number of those collisions, after which a particle leaves the system or freezes out.
This allows us to take that rethermalization\footnote{
The words "immediate rethermalization" used in a few earlier publications, were
badly chosen, misleading and inappropriate.}
happens faster than the freeze out, i.e. that $\lambda_0 < \lambda$ or $\lambda_0\ll \lambda$.
Of course, this argument is true only at the beginning of the FO process, when the density of the
interacting particles is still large.
When $s$ is close to $L$, i.e. near the outer hypersurface,
the first term in eq. (\ref{first-rethermalized}) becomes more important than the rethermalization term
because of its denominator, but as we shall see in the results section, particles freeze out exponentially
fast and for large $s$, when say $99\%$ of the matter is frozen out, the error we introduce with our
approximate solution can not really affect the physical situation.
\\
\begin{figure}[!htb]
\centering
\includegraphics[width=8.6cm, height = 5.6cm]{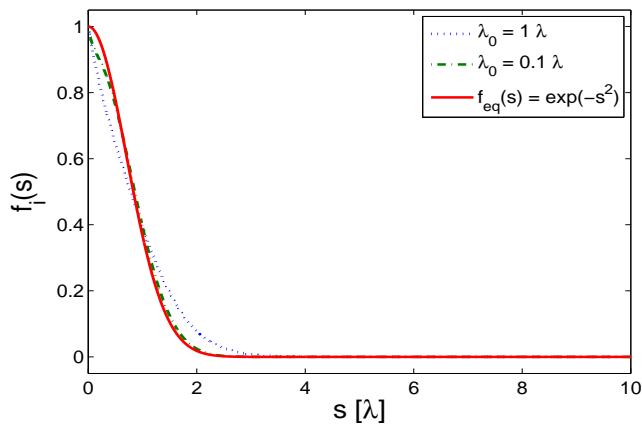}
\caption{
Numerical solutions of eq. (\ref{first-rethermalized}) for different $\lambda_0$'s.
This solution was obtained using, $f_{eq}=e^{-s^2}$, test function, for $L=10\lambda$.
The results show that $f_{i}(s)$ approaches $f_{eq}(s)$ when $\lambda_0\ll \lambda$.
}
\label{solfig}
\end{figure}
\\ \indent
For illustration, let us take a test function, $f_{eq} (s) =  e^{-s^2}$, (we ignore $p$ dependence for the moment),
which is a smoothly and fastly decreasing
function\footnote{In real calculations the $s$-dependence of $f_{eq}(s,p)$ is calculated from
the energy, momentum and baryon charge loss of the interacting component, where these losses are determined
by the momentum dependence of the escape rate and the actual shape of $f_i(s,p)$, as discussed here.
}.
In Fig. \ref{solfig} we show the numerical solutions for the interacting component
for $\lambda_0=\lambda$ and $\lambda_0=0.1\lambda$.
The results show that for the latter case we can safely take for the $f_i$ the approximate
solution \cite{cikk_3,cikk_4,cikk_5}:
\beq
f_{i}(s) = f_{eq}(s)\,.
\eeq{solfi}

\subsection{Conservation Laws}

\begin{figure}[!htb]
\centering
\includegraphics[width=6.8cm, height = 5.2cm]{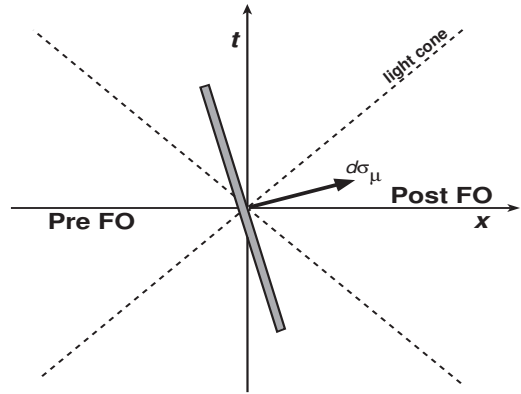}
\caption{ The orientation of the freeze out front in RFG is given as,
$d\sigma_{\mu}=\gamma_{\sigma} (v_{\sigma},1,0,0)_{RFG}$.}\label{FO_frame2}
\end{figure}

\begin{figure}[!htb]
\centering
\includegraphics[width=6.8cm, height = 5.2cm]{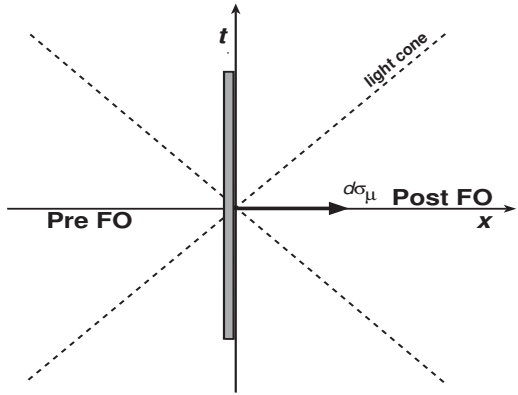}
\caption{ The orientation of the freeze out front in RFF,
i.e. in its own rest frame, is $d\sigma_{\mu} = (0,1,0,0)_{RFF}$.
In this frame the gas has nonvanishing velocity in general.}\label{FO_frame1}
\end{figure}

The goal of the freeze out calculations is to find the final post FO momentum distribution, and then the
corresponding quantities defined through it, starting from the initial pre FO distribution.
On the pre FO side we can have equilibrated matter or gas.
Its local rest frame defines the RFG, see Fig. \ref{FO_frame2}.
We can also define the reference frame, which is attached to the freeze out front,
namely the RFF, see Fig. \ref{FO_frame1}.
These choices are usually advantageous, but other choices are also possible.
\\
\\ \indent
Furthermore, the conservation laws and the nondecreasing entropy condition must be
satisfied \cite{cikk_1}:
\be
\left[ N^{\mu} d\sigma_{\mu}\right] = 0 \, , \quad  \left[ T^{\mu\nu} d\sigma_{\mu}\right] = 0 \, , \quad
\left[ S^{\mu} d\sigma_{\mu}\right] \geq 0 \, ,
\ee
where $ \left[ A \right] = A - A_0$. The pre FO side baryon and entropy currents and the energy-momentum
tensor are denoted by $N^{\mu}_{0}, S^{\mu}_{0}, T^{\mu\nu}_{0}$, while the post FO quantities
are denoted by $N^{\mu},S^{\mu},T^{\mu\nu}$.
\\ \indent
The change of conserved quantities caused by the particle transfer from the interacting matter to the
free matter can be obtained in the following way.
For the conserved  particle 4-current we have
\beq
dN^{\mu}=dN^{\mu}_{i} + dN^{\mu}_{f}=0\ \Rightarrow \  dN^{\mu}_{i} = - dN^{\mu}_{f}\, .
\eeq{dNi}
Then, using the kinetic definition of the particle current, together
with eqs. (\ref{second-rethermalized}, \ref{solfi}), we obtain
\bea \label{dN}
dN^{\mu}_{i} (s)\! &=& -\ ds \!\int \!\frac{d^3 p}{p^0} p^{\mu}
\left[\partial_{s} f_{f} \right] \\ \nonumber
\!&=&\! - \frac{ds}{\lambda} \! \left(\! \frac{L}{L - s} \! \right)\!\!\!
\int \! \frac{d^3 p}{p^0} p^{\mu} \! \left[\frac{ p^s}{p_{\, \rho}u^{\rho}_i}
\Theta(p^s) \right] \! f_{eq}(s) \, .
\eea
Similarly the change in the energy-momentum is
\beq
dT^{\mu\nu}_{i} (s) =  - \frac{ds}{\lambda} \! \left(\!\frac{L}{L - s} \!\right)\!\!\!
\int \! \frac{d^3 p}{p^0}\, p^{\mu} p^{\nu}
\left[\frac{p^s}{p_{\, \rho}u^{\rho}_i}\,
\Theta(p^s)\right] \! f_{eq}(s) \,.
\eeq{dT}
The parameters of the equilibrium (J\"uttner) distribution, $f_{eq}(s)$,
have to be recalculated after each step, $ds$, from the conservation laws as
in \cite{cikk_3,cikk_4,cikk_5}.
\\
\\ \indent
The change of flow velocity, $d u^{\mu}_i(s)$, can be calculated using Eckart's or Landau's definition of
the flow, i.e. from $dN^{\mu}_{i}(s)$  or $dT^{\mu\nu}_{i}(s)$ correspondingly.
Then the change of conserved particle density is given by
\be\label{density}
d n_{i}(s) = u^{\mu}_i(s) \, dN^{\mu}_{i}(s) \, ,
\ee
and for the change of energy density we have
\be\label{energy_density}
d e_{i}(s) = u_{\mu,i}(s) \, dT^{\mu\nu}_{i}(s) \, u_{\nu,i}(s) \,.
\ee
The change of the temperature of interacting component can be found
from this last equation, eq. (\ref{energy_density}), and from the Equation of State (EoS).
This closes our system of equations.
\\ \indent
If we fix the FO direction to the $x$-direction, then eqs. (\ref{dN},\ref{dT}) can be rewritten as:
\bea\label{dNinx}
dN^{\mu}_{i} (x)  &=& -
\frac{dx}{\lambda} \!\left(\! \frac{L}{L-x} \! \right)\! \int \frac{d^3 p}{p^0}\, p^{\mu}\, \\ \nonumber
&\times& \left[\frac{ p\cos \theta_{\vec{p}}}{\gamma(p^0 - jup \, \cos{\theta}_{\vec{p}})}
\Theta(\cos \theta_{\vec{p}}) \right] \! f_{eq}(x,p) \, ,
\eea
and
\bea\label{dTinx}
dT^{\mu\nu}_{i} (x)  &=& -
\frac{dx}{\lambda} \! \left( \! \frac{L}{L-x}\! \right)\! \int \frac{d^3 p}{p^0}\, p^{\mu} p^{\nu}\, \\ \nonumber
&\times& \left[\frac{ p \cos \theta_{\vec{p}}}{\gamma(p^0 - jup \, \cos{\theta}_{\vec{p}})}
\Theta(\cos \theta_{\vec{p}}) \right]\! f_{eq}(x,p) \, ,
\eea
where, the four momentum of particles is $p^{\mu} = (p^0,\vec{p})$, $p=|\vec{p}|$,
$p^x = p \cos \theta_{\vec{p}}$, the flow velocity of the interacting matter is
$u^{\mu}_i = \gamma(1,v,0,0)$, $\gamma = \frac{1}{\sqrt{1 - v^2}}$, $u = |v|$ and $j = sign (v)$ .

\subsection{Changes of the conserved current and energy-momentum tensor}

In this section we show new analytical results for the changes of the conserved particle current and
energy-momentum tensor.
The formulae are analogous to those from ref. \cite{cikk_1,cikk_3}, but now they are calculated with
the Lorentz invariant angular factor from eq. (\ref{esc1}). We show results for both massive and
massless particles.
\begin{widetext}
\bea \nonumber
d N^{0}_i (x) &=& - \frac{dx}{\lambda} \bigg( \frac{L}{L-x} \bigg) \frac{n}{4 u^2 \gamma^2}
\Bigg\{ \gamma G_1^-(m) + \frac{b^3 \Gamma(0,b)}{3 \gamma^2}
- 2\gamma bu\Big[(1+j)K_1(a) - \Kappa_1(a,b)\Big] \\ \nonumber
&-& ub^2 \Big[ (1+j)K_0(a) - \Kappa_0(a,b) \Big]
+\bigg[ u^2 \gamma^2(1 + b)(u^2 - 3) - \frac{A(b)}{3\gamma^2}\bigg]e^{-b} \Bigg\} \\ \nonumber
&\buildrel m=0 \over \longrightarrow & - \frac{dx}{\lambda}
\bigg( \frac{L}{L-x}\bigg) \frac{n}{4} \Bigg[\frac{(3-v)(1+v)^3 }{3} \gamma^2  \Bigg]\,,
\eea
\bea \nonumber
d N^{x}_i (x) &=& \frac{dN^{0}_i (x)}{ju} \, - \,
\frac{dx}{\lambda} \,\bigg( \! \frac{L}{L-x} \bigg) \, \frac{n}{4 u^2 \gamma^2}
\Bigg\{ - 2j \gamma b u^2  \Big[(1+j)K_1(a) - \Kappa_1(a,b)\Big]\\ \nonumber
&-& \, \, \, jb^2 u^2 \Big[(1+j)K_0(a,b) - \Kappa_0(a,b) \Big]
- j u\gamma^2 (1 + u^2) (1+b)e^{-b} \Bigg\}  \\
&\buildrel m=0 \over \longrightarrow & - \frac{dx}{\lambda} \, \bigg(\! \frac{L}{L-x} \bigg)
\frac{n}{4} \Bigg[\frac{2(1 + v)^3}{3}\, \gamma^2  \Bigg]\, ,
\eea
\bea \nonumber
d T^{00}_i (x) &=& - \frac{dx}{\lambda}\, \bigg(\!  \frac{L}{L-x} \bigg) \,
\frac{nT}{4 u^2 \gamma^2}
\Bigg\{ \frac{b^4 \Gamma(0,b)}{4\gamma^3} - \gamma G_2^-(m)
- u \gamma b^2(u^2 + 3)\Big[(1+j)K_2(a) - \Kappa_2(a,b)\Big] \\ \nonumber
 &-& u b^3 \Big[(1+j)K_1(a) - \Kappa_1(a,b) \Big]
+ \gamma^3 \bigg[ - A(b)(3u^2 + 1) + \frac{B(b)}{2 \gamma^6} -
\frac{b^2(3+b)}{3\gamma^6} \\ \nonumber
&-& \frac{b^2u^2(1+b)}{\gamma^6} + \frac{2}{3} b^3 u^4(u^2 - 3) + a^2
\bigg]e^{-b} \, \Bigg\}  \\ \nonumber
&\buildrel m=0 \over\longrightarrow & - \frac{dx}{\lambda}\, \bigg(\! \frac{L}{L-x}\bigg)
\frac{nT}{4}\Bigg[ \frac{(1+v)^4}{2}(6 - 4v + v^2) \gamma^3  \Bigg]\, ,
\eea
\bea \nonumber
d T^{0x}_i (x) &=&  \frac{dT^{00}_i(x)}{ju}\, - \, \frac{dx}{\lambda} \, \bigg( \! \frac{L}{L-x} \bigg) \,
\frac{nT}{4 u^2 \gamma^2} \Bigg\{- j\gamma b^2 (3u^2 + 1)\Big[(1+j)K_2(a) - \Kappa_2(a,b)\Big]\\ \nonumber
&-& j b (b^2 u^2 - 2) \Big[ (1+j)K_1(a) - \Kappa_1(a,b) \Big]
+ j a b\Big[ (1+j)K_0(a) - \Kappa_0(a,b) \Big]  \\ \nonumber &-& ju \gamma \Big[
\gamma^2 A(b) (u^2 + 3)  + (1 + b)(u^2 - 3)
+ \frac{4b^3}{3} u^2 \gamma^2 - b^2 \Big]e^{-b} \Bigg\} \\ \nonumber
&\buildrel m=0 \over \longrightarrow & -
\frac{dx}{\lambda} \, \bigg(\! \frac{L}{L-x}\bigg)\frac{nT}{4} \Bigg[
\frac{(1+v)^4 }{2}(4 - v ) \gamma^3  \Bigg]\, ,
\eea
\bea \nonumber
d T^{xx}_i (x) &=&  \frac{dT^{0x}(x)}{ju} \,- \, 2\frac{T}{\gamma ju} \Bigg[ dN^{x}_i \,- \,
\frac{dN^{0}_i (x)}{ju} \Bigg] \\ \nonumber
&-& \frac{dx}{\lambda} \, \bigg( \! \frac{L}{L-x} \bigg)\, \frac{nT}{4 u^2 \gamma^2}
\Bigg\{ - \gamma u b^2(3 + u^2) \Big[(1+j)K_2(a) - \Kappa_2(a,b)\Big] \\ \nonumber &+&  \gamma^3
\Big[ A(b)(3u^2 + 1)  + \frac{2}{3} b^3 u^4(u^2 - 3) + a^2\Big]e^{-b} \, \Bigg\} \\
&\buildrel m=0 \over \longrightarrow & - \frac{dx}{\lambda} \, \bigg(\! \frac{L}{L-x}\bigg)
\frac{nT}{4} \Bigg[ \frac{3(1+v)^4}{2}
\gamma^3\Bigg]\nonumber\, ,
\eea
\bea \nonumber
d T^{yy}_i (x) &=&  - \frac{dT^{xx}(x)}{2} \, - \,
\frac{dx}{\lambda} \, \bigg(\! \frac{L}{L-x} \bigg) \, \frac{nT}{8 u^2 \gamma^2}
\Bigg\{ - \gamma u b^2(3 + u^2) \Big[(1+j)K_2(a) - \Kappa_2(a,b)\Big] \\ \nonumber
&+& \gamma^3 \Big[ A(b)(3u^2 + 1) - \frac{3 B(b)}{2 \gamma^6} + \frac{(b^2 + 1)(1+b)}{2
\gamma^6} + \frac{2}{3} b^3 u^4(u^2 - 3) + a^2  \Big]e^{-b}
- \frac{b^4 \Gamma(0,b)}{4 \gamma^3} + \gamma H_2^-(m)
\Bigg\} \\ \nonumber
&\buildrel m=0 \over \longrightarrow & - \frac{dx}{\lambda} \, \bigg( \! \frac{L}{L-x} \bigg) \frac{nT}{8}
\Bigg[ \frac{\gamma (1+v)^3}{4 v^2} ( -4 + 12 v - 9 v^2 + 3 v^3) \Bigg] \, ,\nonumber
\eea
and
\be
d T^{zz}_i (x) = d T^{yy}_i(x) \, ,
\ee
\end{widetext}
where, $a = \frac{m}{T}$, $b=a\gamma$, $A(b) = (2 + 2b + b^2)e^{-b}$,
$B(b) =\frac{1}{6}(6 + 6b + 3b^2 + b^3)e^{-b}$ and
$n = 4 \pi T^3 a^2 K_2(a)\, g\, \frac{e^{\mu/T}}{(2 \pi \hbar)^3}$
is the particle density, $g$ is the degeneracy factor while,
$G^-(m)$, $H^-(m)$, $\Kappa(a,b)$, $K(a)$, $\Gamma(0,b)$ are defined in Appendix A.
Note that the $x$-dependent factor $ L/(L-x)$ is just a multiplier in these calculations,
and tends to unity if we are dealing with an infinitely long FO, as in \cite{cikk_1,cikk_3,cikk_4}.

\section{Results and Discussion}

In this section we calculate the post FO distributions and compare the results to former calculations
presented in \cite{cikk_1,cikk_3,cikk_4}.
The effect of two main differences due to the new Lorentz invariant escape rate, eq. (\ref{esc1}), is to be checked:
\begin{enumerate}
\item [-] the infinite, $(\infty)$, FO layer or finite, $(L)$, FO layer,
\item [-] the simple angular escape rate, $P$, or the covariant escape rate, $W$.
\end{enumerate}
We performed calculations for a baryonfree massless gas, where we have used a simple EoS,
$e = \sigma_{SB} T^4$, where $\sigma_{SB} = \frac{\pi^2}{10}$.
The change of temperature is calculated based on this EoS  and eq. (\ref{energy_density}).
There are no conserved charges in our system, consequently we use Landau's definition of the
flow velocity \cite{cikk_3}:
\beq
du^\mu_{i,Landau}(x) = \frac{
\Delta^{\mu\nu}_i(x)\ \ dT_{i,\nu\sigma}(x)\ \ u^\sigma_{i}(x)}{e_i(x) + P_i(x)}\,,
\eeq{dulx}
where
$
\Delta^{\mu\nu}_i(x) = g^{\mu\nu} -  u_{i}^\mu(x)\, u_{i}^\nu(x)\,
$
is a projector to the plane orthogonal to $u^\mu_{i}(x)$, while $e_i(x)$ and $P_i(x)$ are the local
energy density and pressure of the interacting component, i.e.
$T^{\mu\nu}_i(x) =(e_i(x) + P_i(x)) u_{i}^\mu(x) u_{i}^\nu(x)  - P_i(x) g^{\mu \nu}$.
A detailed treatment of Eckart's flow velocity can be found in \cite{cikk_1, cikk_3}.
\\ \indent
For such a system we finally obtain the following set of differential equations:
\bea
d \ln T &=& \frac{\gamma^{2}}{4\sigma_{SB} T^{4}}
\bigg[ dT^{00}_i - 2vdT^{0x}_i + v^{2} dT^{xx}_i \bigg] \, , \\ \nonumber
d v &=&  \frac{3}{4 \sigma_{SB} T^{4}}\bigg[ -v dT^{00}_i + (1+v^{2}) dT^{0x}_i -v dT^{xx}_i \bigg]\, .
\eea
We will present the results for four different cases:
\begin{itemize}
\item [$P_{\infty}$: \,]
We use the simple, but relativistically not invariant
angular factor, $\cos \theta_{\vec{p}}$, in the escape rate,
$$
P_{\infty} = \frac{\cos \theta_{\vec{p}}}{\lambda} \Theta(p \cos \theta_{\vec{p}} ) \, .
$$
The system is characterized by an infinite FO length (up to $x_{max} = 300 \lambda$ in calculations).
The results are shown in Figs. (\ref{1A}, \ref{2A}, \ref{3A}).
This is the same model as in \cite{cikk_1, cikk_3, cikk_4}.
\item [$P_L$: \,\,]
Next, we are using the simple angular factor, but in this case inside a finite FO layer, $L=10 \lambda$,
$$
P_L = \left( \frac{L}{L - x} \right)\frac{\cos \theta_{\vec{p}}}{\lambda} \Theta(p \cos \theta_{\vec{p}} ) \, .
$$
The results are shown in Figs. (\ref{1B}, \ref{2B}, \ref{4A}).
\item [$W_{\infty}$: \,]
Then, we are dealing with the new Lorentz invariant angular factor in the escape rate
and with an infinite FO length, $x_{max} = 300 \lambda$,
$$
W_{\infty}= \frac{1}{\lambda} \left( \frac{p \cos \theta_{\vec{p}}}{p^{\mu} u_{\mu}} \right)
\Theta(p \cos \theta_{\vec{p}} ) \, .
$$
The results are shown in Figs. (\ref{1A}, \ref{2A}, \ref{3A}, \ref{3B}).
\item [$W_L$: \,\,]
Finally we present the primary results of this paper, using both our new improvements, i.e.
the covariant escape rate of eq. (\ref{esc1}),
$$
W_L = \frac{1}{\lambda} \left( \frac{L}{L - x} \right)
\left(\frac{ p \cos \theta_{\vec{p}}}{p^{\mu} u_{\mu}} \right)
\Theta(p \cos \theta_{\vec{p}} )  \,.
$$
The results are shown in Figs. (\ref{1B}, \ref{2B}, \ref{4A}, \ref{4B}).
\end{itemize}
In the case of FO in the infinite layer the factor, $L/(L-x)$, was replaced by $1$.
We presented the situation at a distance of $x_{max} = 300\lambda$, where the amount of still
interacting particles is negligible.
\\ \indent
For the particular cases when we are dealing with an infinite FO, i.e. $P_{\infty}$ and $W_{\infty}$, or
with finite layer FO, i.e. $P_L$ and $W_L$, the results are plotted together.
Thus, in one figure the focus is on the consequences caused by the
different angular factors.
\\ \indent
Thin lines always denote the cases with simple relativistically
not invariant angular factor, these correspond to $P_{\infty}$ and $P_L$.
Thick lines always correspond to cases  with covariant angular factor, $W_{\infty}$
and $W_L$.
\\ \indent
All the figures are presented in the RFF.

\subsection{The evolution of temperature of the interacting component}

\begin{figure}[!htb]
\centering
\includegraphics[width=8.6cm, height = 5.6cm]{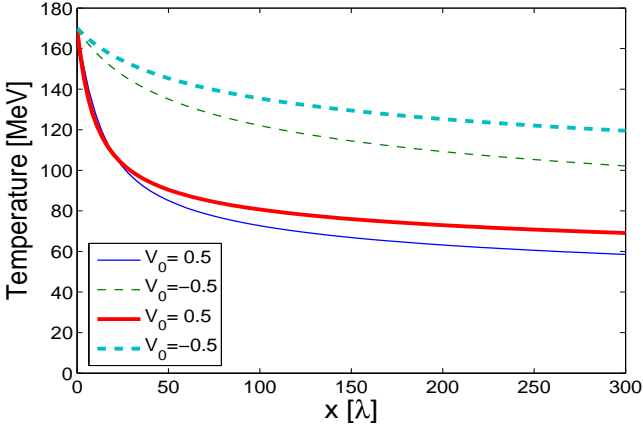}
\caption{The temperature of the interacting component in RFF for a baryonfree massless gas,
calculated with the two escape rates:
$P_{\infty}$ (thin lines) and $W_{\infty}$ (thick lines),
for an infinitely long FO, $(x_{max} = 300\lambda)$.
The initial temperature is \mbox{$T_0=170$ MeV},  $v_0$ is the
initial velocity in RFF.}
\label{1A}
\end{figure}

\begin{figure}[!ht]
\centering
\includegraphics[width=8.6cm, height = 5.6cm]{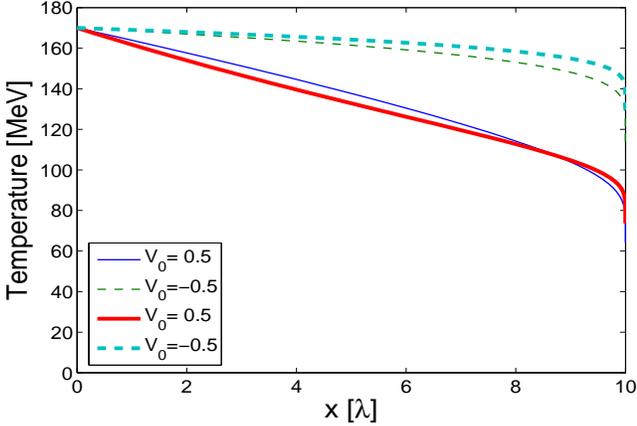}
\caption{The temperature of the interacting component in RFF,
calculated with the
two escape rates: $P_L$ (thin lines) and $W_L$ (thick lines), for a finite, $(L=10\lambda)$, FO layer.
The initial temperature is \mbox{$T_0 = 170\,$ MeV},  $v_0$ is the
initial velocity in RFF.}
\label{1B}
\end{figure}

The first set of figures, Figs. \ref{1A}, \ref{1B}, shows the evolution of temperature of
the interacting component, in fact the gradual cooling of the interacting matter, for the
different cases, $P_{\infty}, P_L, W_{\infty}$ and $ W_L $.
\\ \indent
First, on all figures matter with larger (positive) flow velocity, $v_0$, cools faster.
This is caused by the momentum dependence of the escape rate, which basically tells that
faster particle in the FO direction, will freeze out faster.
Thus, the remaining interacting component cools down, since the most energetic particles freeze
out more often than the slow ones.
Of course, for larger initial flow velocity, $v_0$, in the FO direction, there are more particles
moving in the FO direction with higher momenta in average, than for a smaller flow velocity.
\\ \indent
Now, comparing Fig. \ref{1A} with Fig. \ref{1B}, we can see the difference between finite and
infinite FO dynamics.
In a finite layer the cooling of interacting matter goes increasingly faster as FO proceeds,
while for FO in infinite layer the cooling gradually slows down as $x$ increases.
The reason is the factor, $L/(L-x)$, which speeds up FO as $L-x$ decreases,
and forces it to be completed within $L$.
\\
\\
\\ \indent
The difference  between $P$ and $W$ escape rates comes from the denominator, which
is $p^0$ in case of $P$ and $p^{\mu}u_{\mu}$ in case of $W$.
This difference leads to a stronger cooling for the escape rate $P$ which
is bigger than $W$, if $v_0 \neq 0$.
This can be seen well at later stages of infinitely long  FO, Fig.  \ref{1A}, particularly for the
positive initial flow velocity.
In all other cases the difference between old and new angular factors is insignificant,
what supports our "naive" generalization of the angular factor.

\subsection{The evolution of common flow velocity of the interacting component}

\begin{figure}[!hbt]
\centering
\includegraphics[width=8.6cm, height = 5.6cm]{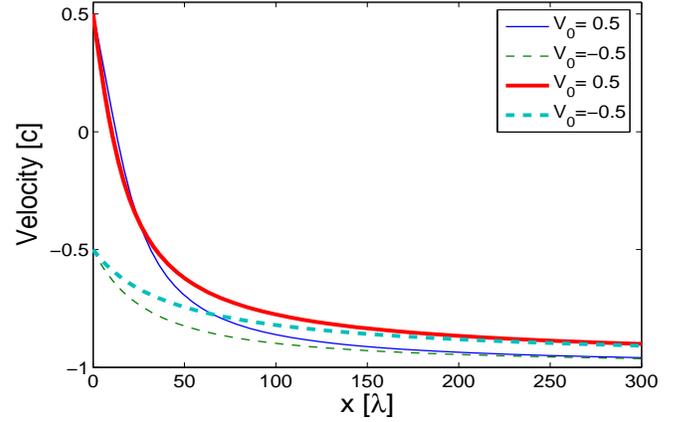}
\caption{The flow velocity of the interacting component in RFF for a baryonfree
massless gas, calculated with the two escape rates: $P_{\infty}$  (thin lines) and  $W_{\infty}$ (thick lines),
for an infinitely long FO, $(x_{max} = 300\lambda)$.
The initial temperature is \mbox{$T_0=170$ MeV}, $v_0$ is the initial velocity in RFF.}
\label{2A}
\end{figure}

\begin{figure}[!hbt]
\centering
\includegraphics[width=8.6cm, height = 5.6cm]{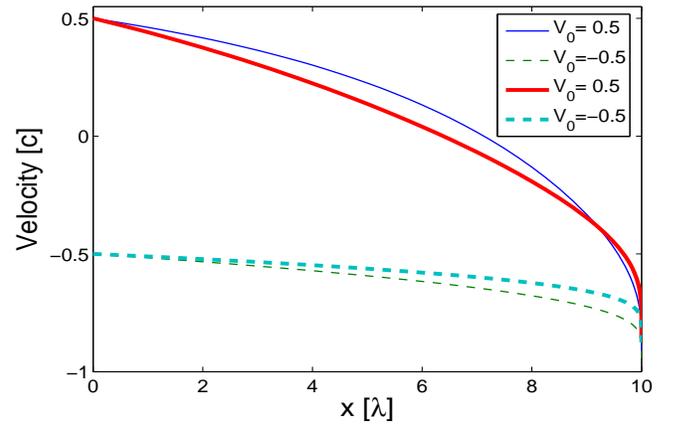}
\caption{The flow velocity of the interacting component in RFF,
calculated with the two escape rates: $P_L$ (thin lines) and $W_L$ (thick lines),
for a finite, $(L=10\lambda)$, FO layer.
The initial temperature is \mbox{$T_0=170$ MeV}, $v_0$ is the initial velocity in RFF.}
\label{2B}
\end{figure}

The second set of figures, Figs. \ref{2A}, \ref{2B}, shows the evolution of the
flow velocity of the interacting component.
\\ \indent
In both cases the flow velocity of the interacting component tends to $-1$, because the FO points
to the positive direction and particles with positive momenta freeze out.
Thus, the mean momentum of the rest must become negative.
\\
\\
\\ \indent
Comparing Fig. \ref{2A} with Fig. \ref{2B}, we can see again that
in a finite layer the flow velocity decreases faster and faster as FO proceeds, while for
FO in infinite layer the velocity change gradually slows down as $x$ increases.
The reason is the $L/(L-x)$ factor, as discussed above.
\\ \indent
The difference between the evolution of the flow velocity, due to the different angular factors,
is again not significant, supporting its generalization.

\subsection{The evolution of the transverse momentum and contour plots of the post FO distribution}

\begin{figure}[!hbt]
\centering
\includegraphics[width=8.6cm, height = 5.6cm]{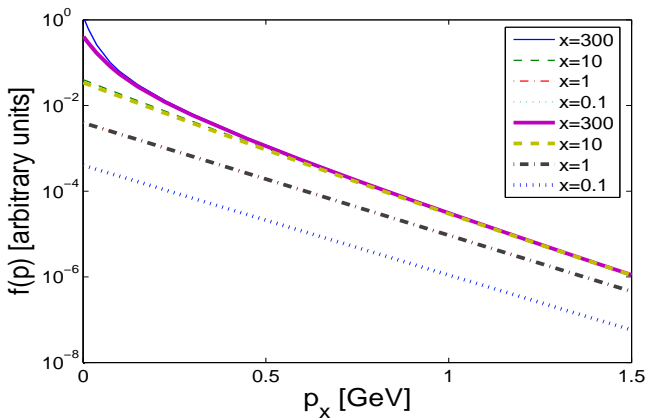}
\caption{The local transverse momentum (here $p_x$) distribution
for a baryonfree massless gas at $(p_y = 0)$,
calculated with the two escape rates: $P_{\infty}$ (thin lines) and $W_{\infty}$ (thick lines),
for an infinitely long FO, $(x_{max}=300\lambda)$.
The initial parameters are $v_0=0$ and $T_0 = 170\,$MeV.
The transverse momentum spectrum is obviously curved due to the freeze out process.
The slope of the transverse momentum distribution increases as we are approaching infinity.}
\label{3A}
\end{figure}

\begin{figure}[!hbt]
\centering
\includegraphics[width=8.6cm, height = 5.6cm]{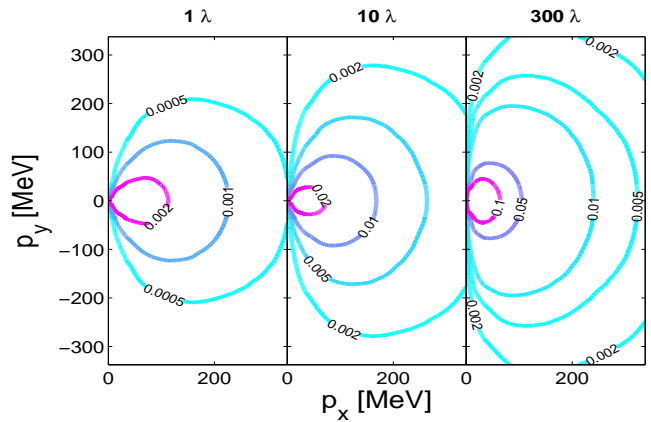}
\caption{The post FO distribution, $f_{f}(x,\vec{p})$, in RFF.
The calculations were done with the Lorentz invariant escape rate, $W_{\infty}$, for an infinitely long FO,
$(x_{max} = 300 \lambda)$.
The subplots correspond to {\bf $x = 1 \lambda,\ 10 \lambda,\  300\lambda$} respectively.
The initial parameters are $v_0=0$ and $T_0 = 170\,$MeV.
Contour lines are given at values represented on the figure.
The maximum is increasing with $x$ as indicated in Fig. \ref{3A}.
The distribution is asymmetric and elongated in the FO direction.  This may lead
to a large-$p_t$ enhancement, compared to the usual  J\"uttner
assumption used in many earlier calculations as a post freeze out distribution.
Note that $f_{f}(x,\vec{p})$ does not tend to the cut J\"uttner
distribution even at very large $x$.}
\label{3B}
\end{figure}

\begin{figure}[!hbt]
\centering
\includegraphics[width=8.6cm, height = 5.6cm]{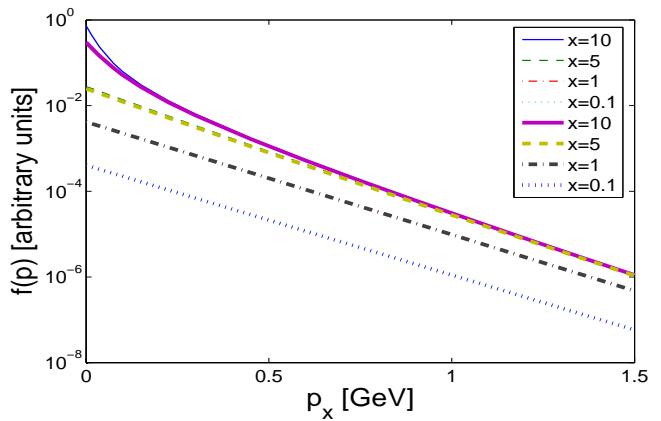}
\caption{The local transverse momentum (here $p_x$) distribution for a baryonfree
massless gas at $(p_y = 0)$, calculated with the two escape rates:
$P_L$ (thin lines) and  $W_L$
(thick lines) for a finite, $(L=10\lambda)$, FO layer.
The initial parameters are $v_0=0$ and $T_0 = 170\,$MeV.}
\label{4A}
\end{figure}

\begin{figure}[!hbt]
\centering
\includegraphics[width=8.6cm, height = 5.6cm]{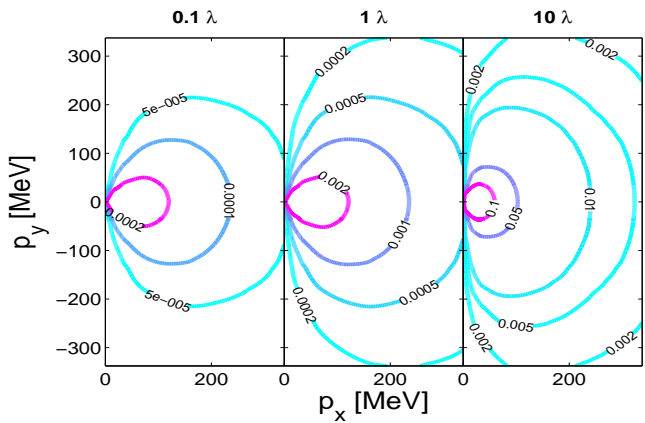}
\caption{The post FO distribution, $f_{f}(x,\vec{p})$, in RFF.
The calculation was made using the covariant escape rate, $W_L$ for a finite, $(L=10\lambda)$, FO layer.
The subplots correspond to {\bf $x = 0.1 \lambda,\ 1 \lambda,\  10\lambda$} respectively.
The initial parameters are $v_0=0$ and $T_0 = 170\,$MeV.
Contour lines are given at values represented on the figure.}
\label{4B}
\end{figure}

The next set of figures, Figs. \ref{3A},  \ref{3B}, shows the evolution of the transverse
momentum distribution, while Figs. \ref{4A}, \ref{4B}, present the contour plots
of the post FO momentum distribution, for $W_{\infty}$ and $W_{L}$.
We have presented a one-dimensional model here, but we assume that it is applicable for the
direction transverse to the beam in heavy ion experiments.
The presented plots should be qualitatively compared to the transverse momentum
distributions of measured pions.
\\
\\ \indent
What we see is that all the final post FO momentum distributions are essentially the same.
This is very important outcome from our analysis, which we will discuss below.
Also, one can see that resulting post FO distributions are non-thermal distributions,
as it has been shown already in \cite{cikk_1,cikk_3,cikk_4},
they strongly deviate from exponential form in the low momentum region.
The increase in the final FO spectra over the thermal distribution
for low momenta is connected to the fact that at late stages of the FO process,
the interacting component is cold and its flow velocity is negative.
So, it contributes only to the low momentum region of the post FO spectra.
\\
\\ \indent
These results were obtained in a stationary one-dimensional model with a single flow velocity.
In reality different space-time sections of the overall FO layer are moving with respect to each other
with considerable velocities, i.e. $v\approx 0.2 - 0.7$.
Therefore, the superposition of these parts of the FO layer wash out the very sharp peaks at small
momenta, while the curvature at higher momenta, although it is smaller, may persist even after superposition.
There are several effects mentioned in the literature, which can cause such a curvature.
The effects discussed in this section, arising from kinetic description, may contribute to the
curvature of the spectra, but we need a more realistic full scale, nonstationary 3-dimensional model
to estimate the expected shape of the $p_t$ spectra in measurements.
Consequently, both the contributions of space-like and time-like sections of the FO layer have to contribute.
\\

\subsection{Freeze out in layers of different thickness}

\begin{figure}[!hb]
\centering
\includegraphics[width=8.6cm, height = 5.6cm]{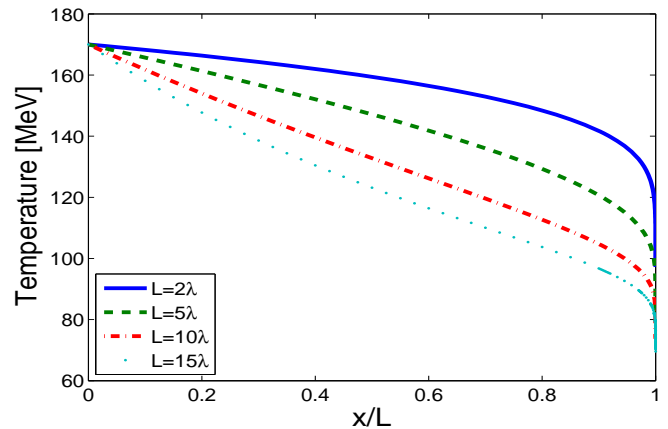}
\caption{The evolution of the temperature of the interacting component in RFF,
for a baryonfree massless gas, calculated with the $W_L$ escape rate for
different FO layer thicknesses $L=2\lambda,\ 5\lambda,\ 10\lambda,\ 15\lambda$.
The initial parameters are $v_0=0.5$ and $T_0 = 170\,$MeV. }
\label{fig1M1}
\end{figure}

\begin{figure}[!hbt]
\centering
\includegraphics[width=8.6cm, height = 5.6cm]{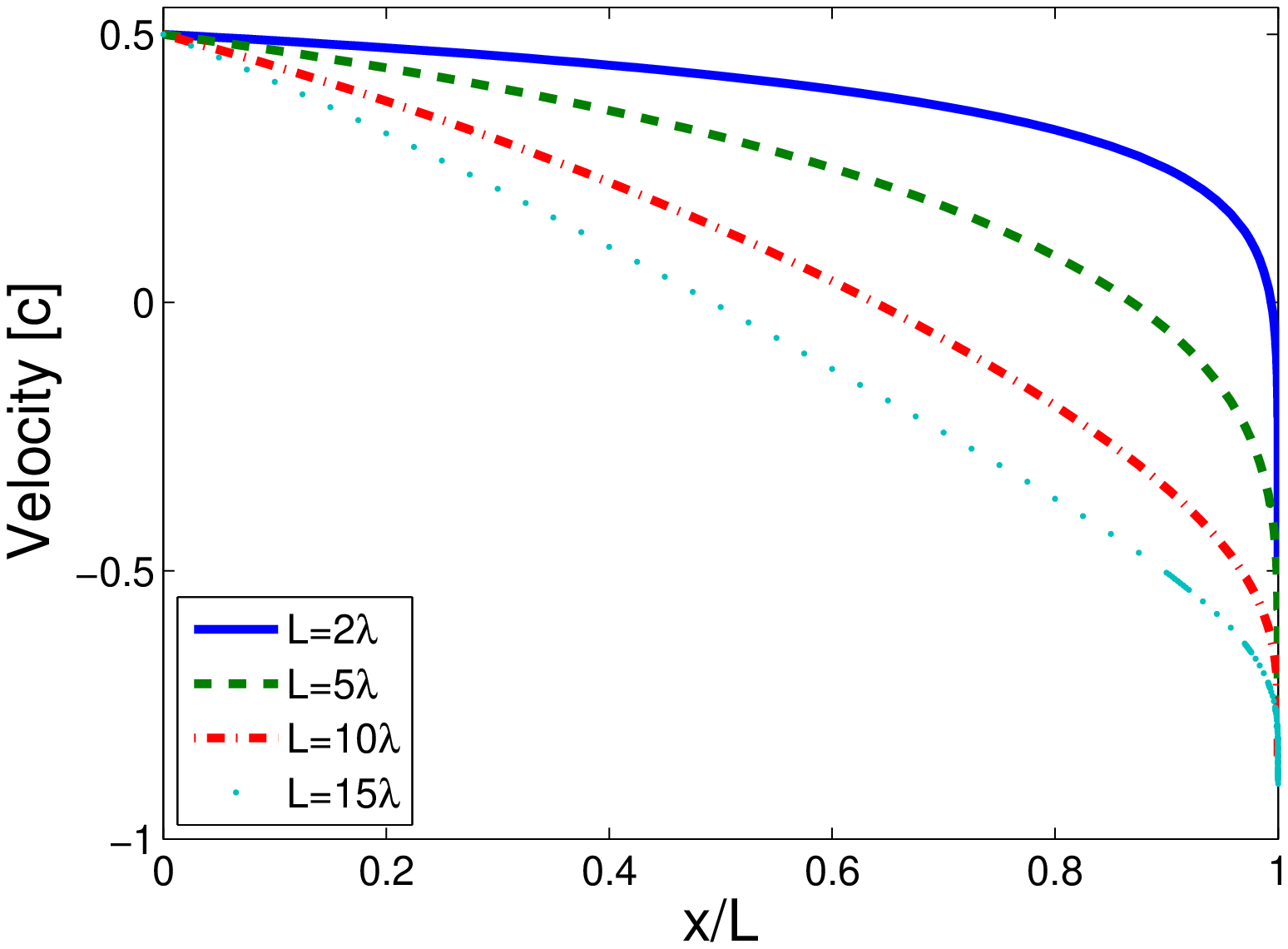}
\caption{The evolution of the flow velocity of the interacting component in RFF,  for a baryonfree
massless gas,
calculated with the $W_L$ escape rate  for different FO layer thicknesses
$L=2\lambda,\ 5\lambda,\ 10\lambda,\ 15\lambda$.
The initial parameters are $v_0=0.5$ and $T_0 = 170\,$MeV.  }
\label{fig1M2}
\end{figure}

\begin{figure}[!htb]
\centering
\includegraphics[width=8.6cm, height = 5.6cm]{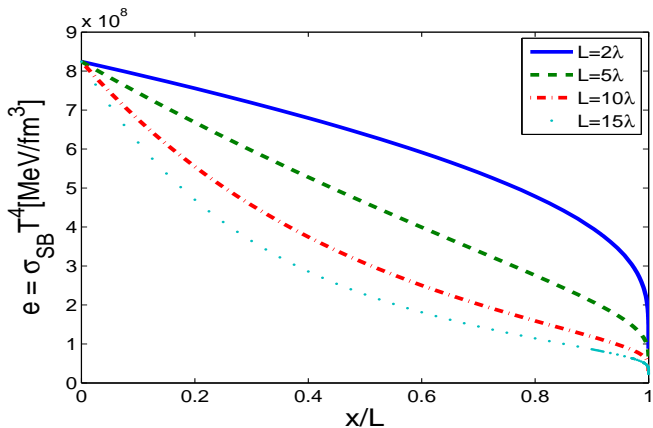}
\caption{The evolution of the energy density of the interacting component in RFF,  for a baryonfree
massless gas, $\sigma_{SB}=\pi^2/10$,
calculated with the $W_L$ escape rate  for different FO layer thicknesses
$L=2\lambda,\ 5\lambda,\ 10\lambda,\ 15\lambda$.
The initial parameters are $v_0=0.5$ and $T_0 = 170\,$MeV. }
\label{ener}
\end{figure}

\begin{figure*}[!htb]
\centering
\includegraphics[width=17cm, height=8cm]{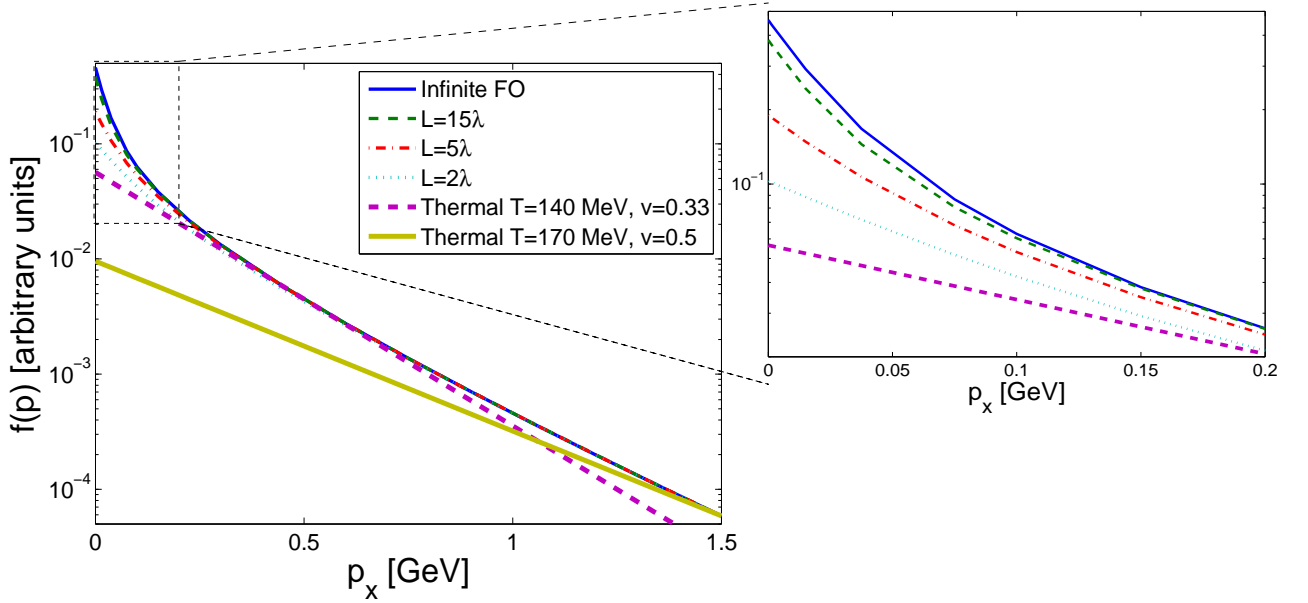}
\caption{Final post FO transverse momentum (here $p_x$)  distributions for the FO layers of different thicknesses.
Calculations were done for baryonfree massless gas with escape rate $W_L$ for $L=2\lambda,\ 5\lambda,\ 15\lambda$
and with $W_{\infty}$ (thick line).
The initial conditions are the same as in Figs. \ref{fig1M1}, \ref{fig1M2}: $T_0=170\ MeV$, $v_0=0.5$.
Distributions for the different FO layer thicknesses are very similar, with some difference
in the low momenta region, which is shown in more detail in the "zoom in" subplot.
The two thick lines correspond to some effective thermal distributions, with the corresponding
parameters displayed in the plot legend.
These are shown to illustrate the difference between obtained post FO distributions
and thermal distributions.
}
\label{fig2M}
\end{figure*}

In this section we show the results of calculations performed with $W_L$ escape rate, for different
finite FO layer thicknesses.
Some results of such an analysis have also been presented in \cite{QM05}.
\\ \indent
In Figs. \ref{fig1M1}, \ref{fig1M2}, we present the evolution of the temperature and flow velocity of
the interacting component for $L=2\lambda,\ 5\lambda,\ 10\lambda,\ 15\lambda$.
We plot the resulting curves as function of $x/L$, what allows us to present them all in one figure.
We clearly see, and this agrees also with our previous comparison to infinitely long FO, that
by introducing and varying the thickness of the FO layer, we are strongly affecting the evolution
of the interacting component.
\\ \indent
We can also study how fast the energy density of the interacting component is decreasing, see Fig. \ref{ener}.
Since there is no expansion in our simple model, the evolution of the energy density is
equivalent to the evolution of the total energy of the remaining interacting matter.
We can see that the decrease of the energy density of the interacting component is exponentially fast,
what justifies our way of getting approximate an solution for the interacting component,
see section \ref{Kinetic model}.
\\
\\
\\
\\ \indent
Figure \ref{fig2M} shows the final post FO transverse momentum distribution for different $L$.
Despite the differences in the evolution of the interacting component,
all the final post FO distributions look the same and are practically indistinguishable.
The difference between the result for a FO layer as thin as $L=2\lambda$ and that for
$L\rightarrow \infty$ limit shows up only in the low momentum region, and it is not
significant enough to allow us to resolve layers of different thicknesses
from experimental spectra.
Thus, the thickness of the FO layer does not affect, as we have seen already in the previous section,
the final post FO distribution, which is in fact the measured quantity!

\section{Conclusions and outlook}

In this paper we presented a simplified, but still non-trivial, Lorentz invariant freeze out model,
which allows us to obtain analytical results in the case of a massless baryonfree gas.
In addition the model realizes freeze out within a finite freeze out layer.
\\ \indent
We do not aim to apply directly the results presented here to the experimental heavy ion collision data,
instead our purpose was to study qualitatively the basic features of the freeze out effect, and
to demonstrate the applicability of this covariant formulation for FO in finite length.
\\ \indent
In Figs. \ref{1A}, \ref{2A}, \ref{3A} and Figs. \ref{1B}, \ref{2B}, \ref{4A},
we compare results, with the simple, $\cos \theta_{\vec{p}}$, angular factor,
and with  the Lorentz invariant angular factor, $\frac {p^\mu d\sigma_\mu}{p^\mu u_\mu}$.
The differences are insignificant, supporting our generalization.
\\ \indent
As it has been indicated in the previous publications \cite{cikk_1, cikk_3, cikk_4},
the final post FO distributions are non-equilibrated distributions,
which deviate from thermal ones particularly in the low momentum region.
The final spectra have a complicated form and were calculated here numerically.
In large scale (e.g. 3-dim CFD) simulations for space-like FO the
Cancelling J\"uttner distribution \cite{karolis}, may be a satisfactory analytical approximation.
\\ \indent
Our analysis shows that by introducing  and varying the thickness of the FO layer,
we are strongly affecting the evolution of the interacting component,
but the final post FO distributions, even for small thicknesses, e.g. $L=2\lambda$,
look very close to our results for an infinitely long FO, first obtained in \cite{cikk_1, cikk_3, cikk_4}.
\\ \indent
The results suggest that if the measured post FO spectrum is curved, as shown in Fig. \ref{fig2M},
then it doesn't matter how thick FO layer was, and we do not need to model the details of
FO dynamics in simulations of collisions!
Once we have a good parametrization of the post FO spectrum (asymmetric, non-thermal),
it is enough to write down the conservation laws and non-decreasing entropy
condition with this distribution function \cite{cikk_2}, (and probably with some volume scaling
factor to effectively account for the expansion during FO).
This Cooper-Frye type of description can be viewed from two sides.
From experimental side, when we know the post FO spectra, we can extract information about the conditions
in the interacting matter before FO.
In theoretical, e.g. fluid dynamical, simulations such a procedure would allow us to calculate parameters
of the final post FO distributions to be compared with data.
In this way our results may justify the use of FO hypersurface in hydrodynamical models for heavy ion collisions,
but with  proper non-thermal post FO distributions.
\\ \indent
At the same time, while the final distribution, $f(p)$, is not sensitive to the kinetic evolution, other
measurables, especially the two particle correlation function may be more sensitive to the details
and extent of the FO process.
\\ \indent
The model can also be applied to FO across a layer with time-like normal.
While several of the conclusions can be extended to the time-like case, it also requires
still additional studies \cite{article_2}.
\\ \indent
For realistic simulations of high energy heavy-ion reactions the full
3D description of expansion and FO of the system should be modeled simultaneously.
We believe that our invariant escape rate, can be a basic ingredient of such models.

\begin{acknowledgments}

One of the authors, L. P. Csernai, thanks the
Alexander von Humboldt Foundation for extended support in
continuation of his earlier Research Award. The authors thank the
hospitality of the Frankfurt Institute of Advanced Studies and the
Institute for Theoretical Physics of the University of Frankfurt, and
the Gesellschaft f\"ur Schwerionenforschung, where parts of this
work were done.
\\ \indent
E. Moln\'ar, thanks the hospitality of Justus - Liebeg University of Giessen, where
parts of this work were done under contract number, HPTM-CT-2001-00223, supported by
the EU-Marie Curie Training Site.
\\ \indent
Enlightening discussions with Cs. Anderlik, Zs. I. L\'az\'ar and T. S. Bir\'o are
gratefully acknowledged.
\end{acknowledgments}
\appendix
\section{}
The definition of $\Kappa_{n}(z,w)$ function is:
\be
\Kappa_{n}(z,w) = \frac{2^n \, n!}{(2n)!} \, z^{-n}
\int_{w}^{\infty} dx \,e^{-x} \,(x^2 - z^2)^{n-\frac{1}{2}} \, ,
\ee
where in the case of $w=z$ and $n>-1$ the above formula will lead to the
modified Bessel function of second kind, $K_{n}(z)$.
Furthermore, the indefinite integral \cite{mathematica} is:
\be
\int z^{\alpha - 1} \, \Gamma(n,z) dz = \frac{z^{\alpha} \, \Gamma(n,z) - \Gamma(n + \alpha,z)}{\alpha} \, ,
\ee
where $\Gamma (n,z)$ is the incomplete gamma function:
\be
\Gamma(n,z) = \int_{z}^{\infty} dt \, t^{n-1} \,e^{-t} \, .
\ee
The analytically not integrable functions $G_n^- (m) $ and $H_n^-(m)$ are defined as:
\bea G_n^-(m)
&=&\frac{1}{T^{n+2}}\int_{0}^{\infty} d p \, p \, \left( \sqrt{p^2 + m^2} \right)^n \, \\ \nonumber
&\times&\Gamma \Big(0,\frac{\gamma}{T} \sqrt{p^2 + m^2} - \frac{\gamma jup}{T}\Big) \, ,
\eea
and
\bea
H_n^-(m) &=&\frac{1}{T^{n+2}}\int_{0}^{\infty} d p \, p^{n+1} \, \\ \nonumber
&\times& \Gamma \left(0,\frac{\gamma}{T} \sqrt{p^2 + m^2} -
\frac{\gamma jup}{T}\right) \, .
\eea
\\ \indent
In the massless limit, we have the $G_n^- (0) = H_n^- (0)$.
Values of these functions for  $(n = 1,2)$ are given below:
\bea
G_1^-(0) &=& \frac{1}{T^3} \,
\int_{0}^{\infty} d p \, p^2  \, \Gamma \left(0,\frac{\gamma}{T}\, p
( 1 - ju)\right) \\ \nonumber
&=& \frac{2 }{3 \gamma^3}( 1 - ju)^{-3} \, , \\
G_2^-(0) &=& \frac{1}{T^4} \, \int_{0}^{\infty} d p \, p^{3}  \,
\Gamma \left(0,\frac{\gamma}{T}\, p ( 1 - ju ) \right) \\ \nonumber
&=& \frac{3 }{2 \gamma^4}( 1 - ju)^{-4} \, .
\eea
In the general calculation of the integrals in RFF, we change variables from $p$ to $z$ as given below:
\bea
&\!& \int_{0}^{\infty}  dp \, f(p) \, e^{-\frac{\gamma}{T}(\sqrt{p^2 + m^2} - jup)}  \\ \nonumber
&=& j\gamma T \int_{b}^{a} dz \Big( u -  \frac{z}{\sqrt{z^2 - a^2}} \Big)\,e^{-z} \\ \nonumber
&\times& f \Big[\gamma T(juz - j \sqrt{z^2 - a^2} )\Big]  \\\nonumber
&+& j\gamma T  \int_{a}^{\infty} dz \Big( u + j \frac{z}{\sqrt{z^2 - a^2}} \Big) \,  e^{-z} \\ \nonumber
&\times& f \Big[\gamma T(juz + \sqrt{z^2 - a^2} )\Big] \, , \\ \nonumber
\eea
where $z = \gamma(\sqrt{p^2 + m^2} - jup)/T$, $a=m/T$ and $b=\gamma a$.



\end{document}